\documentclass{PoS}

\title{The high voltage system for the novel MPGD-based photon detectors of COMPASS RICH-1}

\ShortTitle{HV system for MPGD-based photon detectors}

\author{J. Agarwala$^a$,
R. Birsa$^a$,
F. Bradamante$^{a,b}$,
A. Bressan$^{a,b}$,
C. Chatterjee$^{a,b}$,
P. Ciliberti$^{a,b}$,
\speaker{S.~Dalla~Torre}$^a$\thanks{corresponding author},
S. Dasgupta$^a$,
B. Gobbo$^a$,
M. Gregori$^a$,
G. Hamar$^a$,
S. Levorato$^a$,
A. Martin$^{a,b}$,
G. Menon$^a$,
F. Tessarotto$^a$,
Y. Zhao$^a$
\\
\llap{$^a$}INFN, Sezione di Trieste, 
Trieste, Italy\\
\llap{$^b$}University of Trieste,  
Trieste, Italy\\
E-mail: \email{Silvia.DallaTorre@ts.infn.it}\\
}
\abstract{The architecture of the novel MPGD-based photon detectors of COMPASS RICH-1 consists in a large-size hybrid MPGD multilayer layout combining two layers of Thick-GEMs and a bulk resistive MICROMEGAS. Concerning biasing voltage, the Thick-GEMs are segmented in order to reduce the energy released in case of occasional discharges, while the MICROMEGAS anode is segmented in pads individually biased at positive  voltage, while the micromesh is grounded. In total, there are ten different electrode types and more than 20000 electrodes supplied by more than 100 HV channels. Commercial power supply units are used.  The original elements of the power supply system are the architecture of the voltage distribution net, the compensation, by voltage adjustment, of the effects of pressure and temperature variation affecting the detector gain and a sophisticated control software, which allows to protect the detectors against errors by the operator, to monitor and log voltages and current at 1~Hz rate and to automatically react to detector misbehaviors.
\par
The HV system and its performance are described in detail as well as the electrical stability of the detector during the operation at COMPASS.
}

\FullConference{5th International Conference on Micro-Pattern Gas Detectors (MPGD2017)\\
		22-26 May, 2017\\
		Philadelphia, USA}
\begin{document}
\section{Introduction}
Novel single photon detectors based on MPGD 
technologies~\cite{upgrade} have
been developed for the upgrade
of the RICH-1 detector~\cite{rich1} of
the COMPASS~\cite{compass} experiment at
CERN SPS. They have been installed during the 2015-2016
Winter shut-down of the SPS and operated during 
the 2016 and 2017 COMPASS data taking. 
\section{Requirements}
\begin{figure}
\begin{center}
\includegraphics[width=0.6\textwidth]{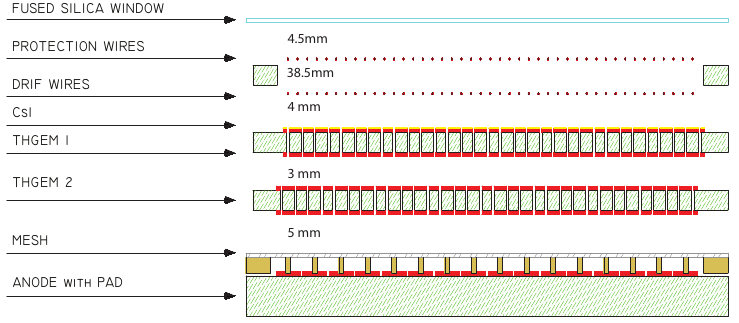}
\end{center}
\caption{\label{fig:hybrid}
Sketch of the hybrid single photon detector (image not to scale).}
\end{figure}

\begin{figure}
\begin{center}
\includegraphics[width=\textwidth]{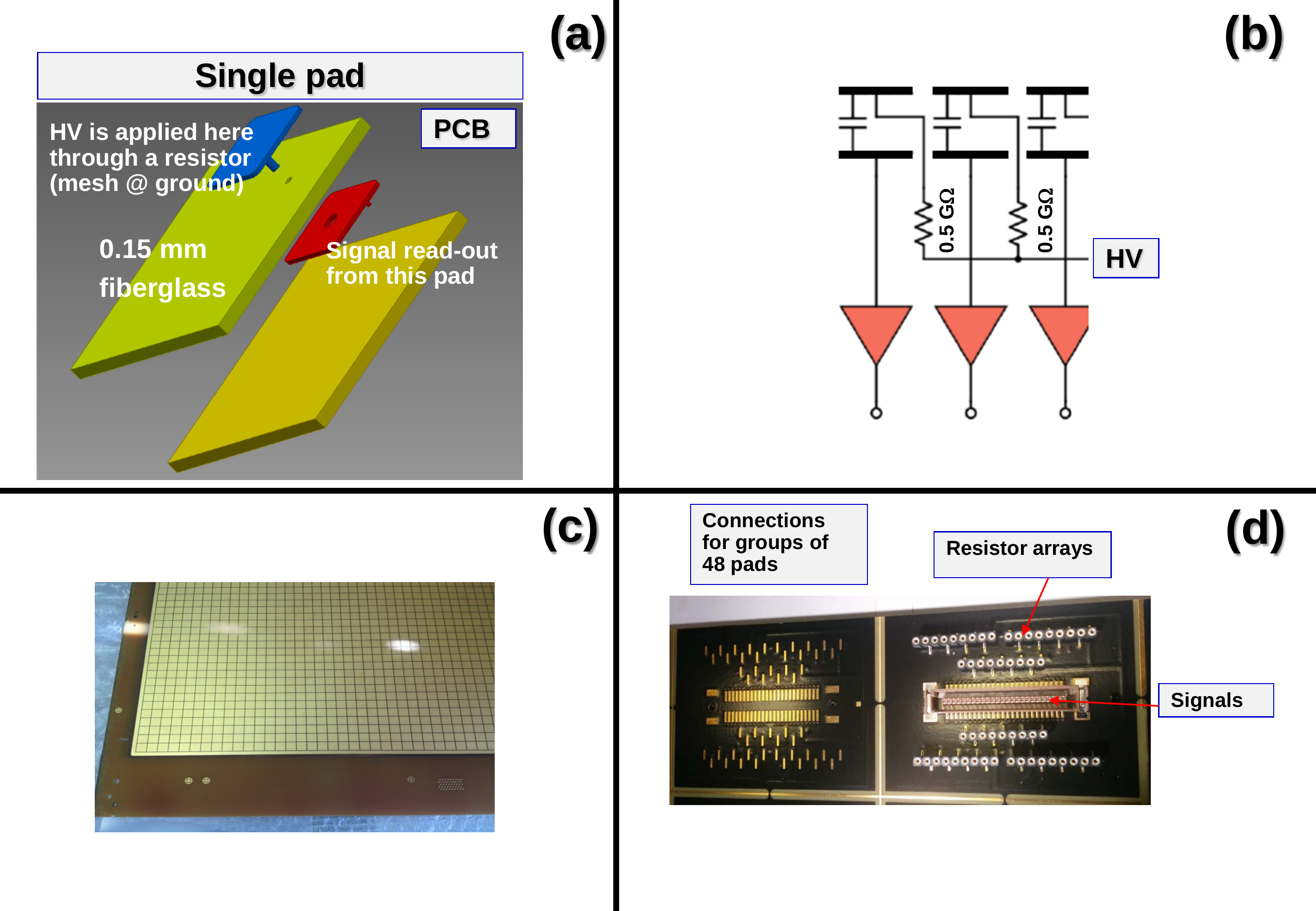}
\end{center}
\caption{\label{fig:resistiveMM}
The resistive MM by discrete elements. (a) Sketch illustrating the principle by the exploded view of the PCB layers: the blue pad is the anode electrode of the MM; the red pad is embedded in the PCB and the signal is transferred from the blue to the red pad by capacitive coupling. (b) The principle is illustrated by the electrical scheme: the top elements of the capacitors are the pad forming the MM anode (blue pad in (a)), the bottom elements of the capacitors (red pad in (a)) are connected to the front-end electronics. (c) Picture of the MM anode PCB, front view. (d) Picture of the MM anode PCB, rear view, detail: the connectors serving 48 pads are grouped together; the signal connectors and the resistor arrays connectors can be seen.}
\end{figure}

The requirements for the High Voltage
(HV) system 
originate from the architecture of the
novel detectors,  consisting
in a hybrid MPGD arrangement 
(Fig.~\ref{fig:hybrid}): 
two layers of THick GEMs
(THGEM)~\cite{thgem}, the first one 
also acts as a reflective
photocathode (its top face is coated 
with a CsI film), are
coupled to a MicroMegas (MM)~\cite{mm} 
on a pad segmented
anode. The MM is resistive by an 
original implementation making use of discrete elements: 
HV is applied to the anode 
pads, each one  protected 
by an individual resistor, while the 
signals are collected from a second 
set of pads, parallel to the first ones, 
embedded in the anode PCB 
where the signal is transferred by 
capacitively coupling 
(Fig.~\ref{fig:resistiveMM}). 
The detectors are operated 
with a gas mixture Ar~:~CH$_4$~=~50~:~50, selected for 
optimal extraction of the photoelectrons from the 
converting CsI film to the gaseous atmosphere~\cite{rich1}.
\par
\begin{figure}
  \begin{minipage}[b]{0.4\textwidth}
\includegraphics[width=\textwidth]{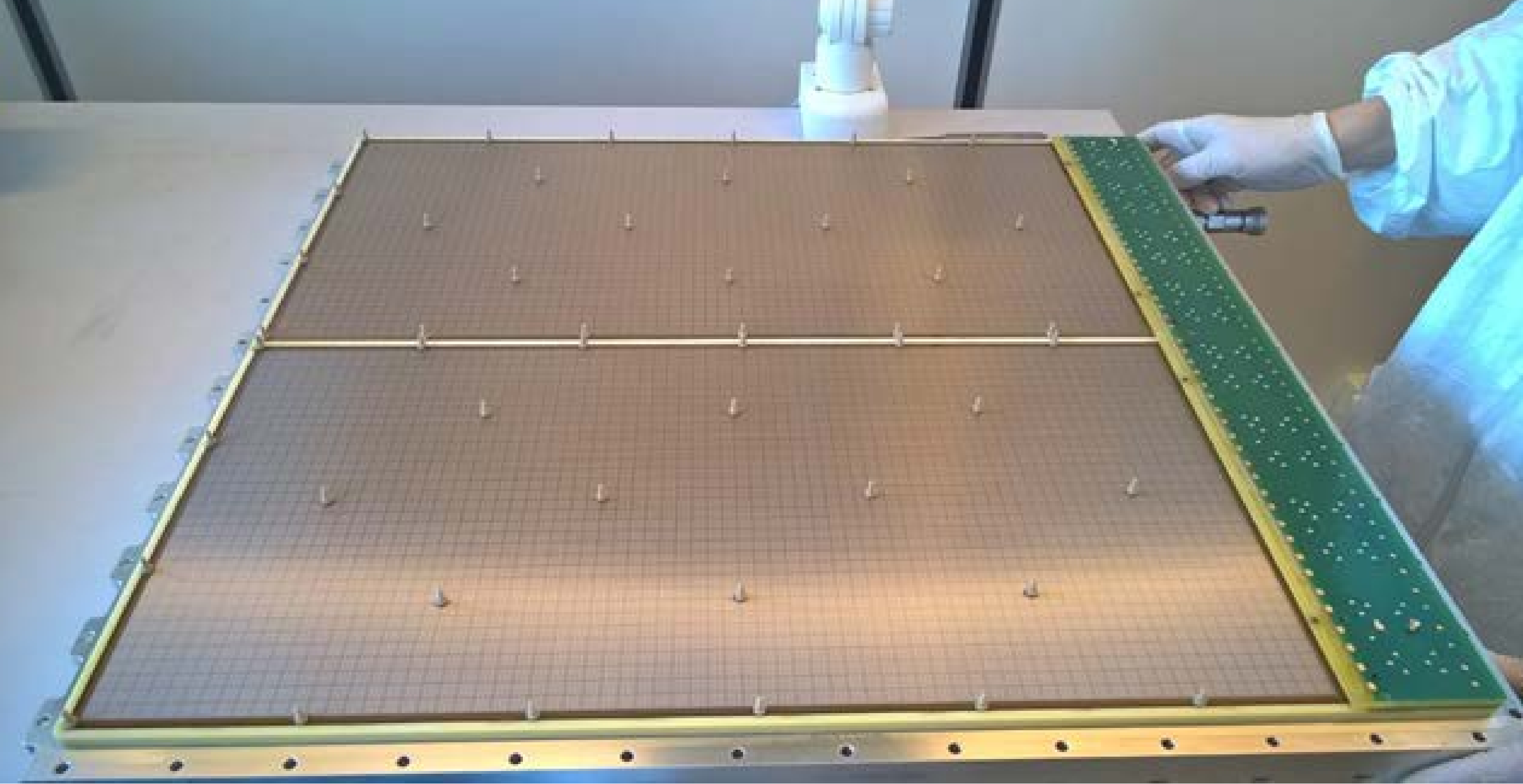}    \caption{\label{fig:mm-picture}
    Picture, taken during the detector construction,
of two MM units arranged in a detector.}
  \end{minipage}
  \hfill
\begin{minipage}[b]{0.4\textwidth}
\includegraphics[width=\textwidth]{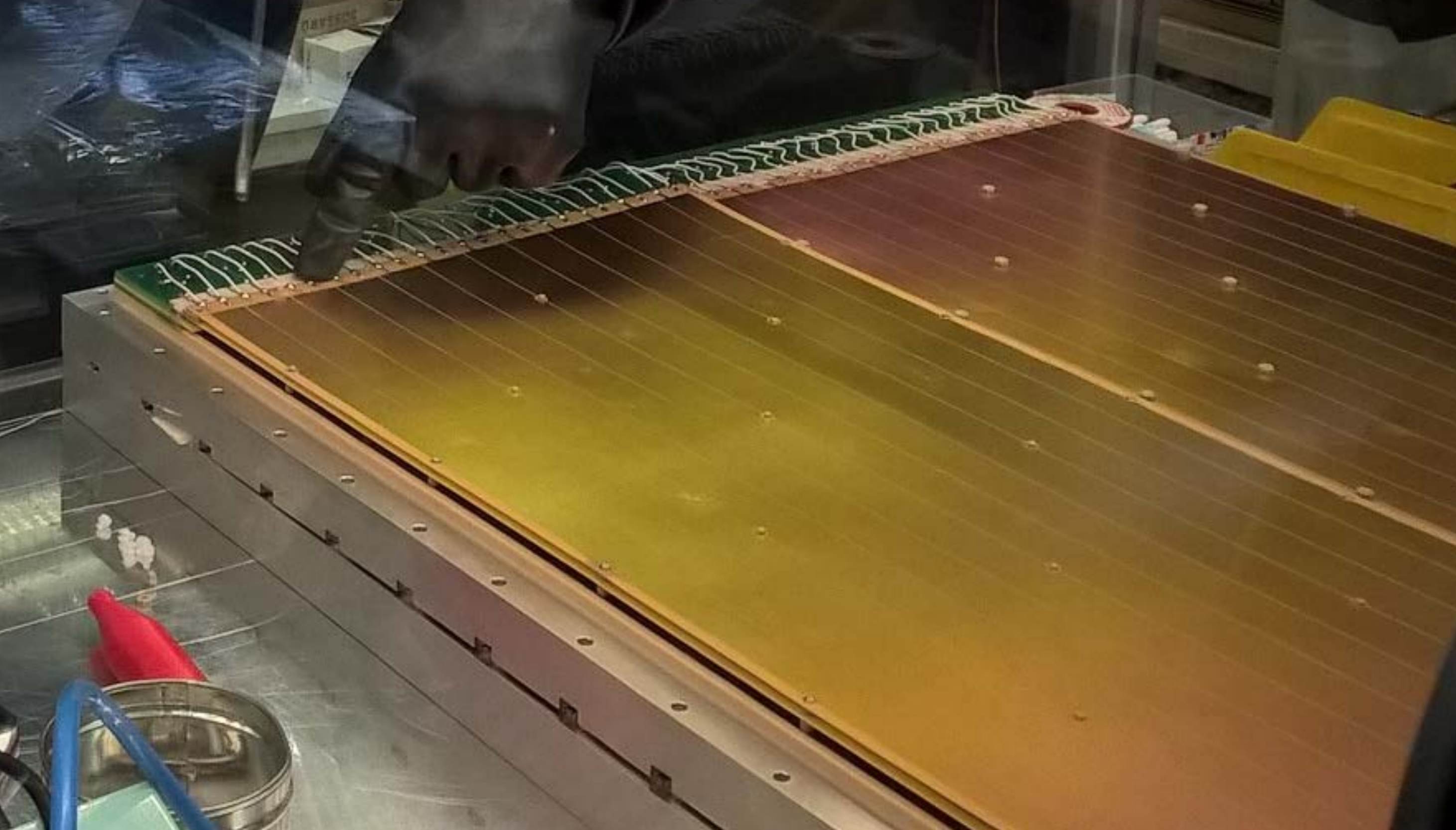}
\caption{\label{fig:detector-thgem-picture}
Picture, taken during the detector construction,
of two THGEMs arranged in a detector. The 12 
physical segments of each THGEM are visible.}    
\end{minipage}
\end{figure}

For what concerns the active multipliers, namely THGEMs and
MM gaps, each 600$\times$600~mm$^2$ detector is formed by two 
600$\times$300~mm$^2$ units arranged side 
by side within a single 
detector (Fig.s~\ref{fig:mm-picture}, 
\ref{fig:detector-thgem-picture}), while the
protection and drift wire planes are in 
common for the two units.
\par
The typical voltages applied to the detector electrodes are 
(with reference to electrodes as indicated in 
Fig.~\ref{fig:hybrid}): protection wires at -300~V, 
drift wires at -3400~V, THGEM~1, top at -3200~V, 
THGEM~1, bottom at -2000~V,
THGEM~2, top at -1700~V, THGEM~2, bottom at -500~V, 
mesh grounded
and anode at +600~V.
\section{The high voltage distribution scheme and the high voltage system}
The protection and drift wires are powered by a
power supply channel each.
\par
\begin{figure}
\begin{center}
\includegraphics[width=0.4\textwidth]{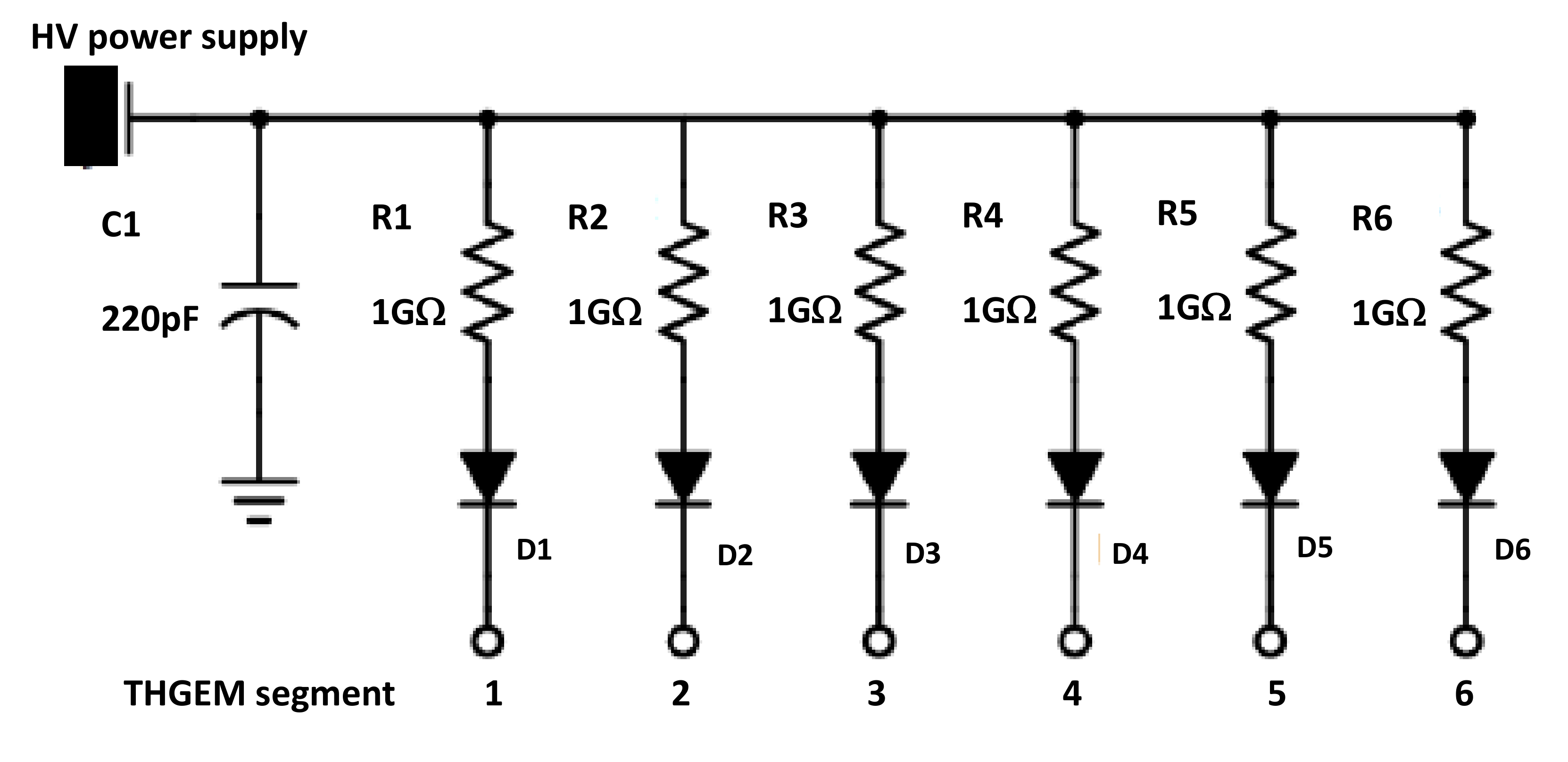}
\end{center}
\caption{\label{fig:thgem-hv}
Scheme of the voltage distribution to six THGEM segments of the top (bottom) face.}
\end{figure}

The THGEM plates are physically segmented 
in 12 longitudinal segments 
with area 24$\times$600~mm$^2$ 
(Fig.~\ref{fig:detector-thgem-picture}).
Concerning power supply, six segments of the
top (bottom) face, corresponding to an area of 
180$\times$600~mm$^2$ referred to as a "sector" 
in the following, are powered 
by a single individual supply 
channel, where the voltage is distributed 
to the individual segments according 
to the scheme shown in Fig.~\ref{fig:thgem-hv}. 
The resistors and fast diodes in series with each 
segment aim at suppressing the cross-talk of the 
voltage swing when a discharge affects one of the segments.
\par
The distribution scheme of the power to the MM pads 
is shown in Fig.~\ref{fig:resistiveMM}~(b) 
and two power supply channels 
per unit are used, each one supplying 2450 pads;
this area corresponds to that of a sector:
from the power supply point of view, each detector, 
formed by two units, is powered by sectors, 
two per unit and, therefore,
four per chamber.
\par
The preservation of the correct electric field configuration 
at the lateral detector edges, where the detector frames are 
very near to the active elements in order to minimize 
the dead area between adjacent detectors,  imposes 
to preserve the field shape by
the presence of two auxiliary 
electrodes at the two lateral sides of the detector frames, 
corresponding to further four supply channels per detector.
\par
In total, the supply system for the four detector set includes 
more than 100 channels. Commercial power supply units 
by CAEN~\footnote{CAEN S.p.A., 
Via della Vetraia, 11, 55049 Viareggio, Lucca, Italy} 
are used.
THGEMs are powered by 
A1561HDN 12 channel units, capable of voltage supply up to  
-6~kV and current monitor with 50~pA resolution, fully 
satisfactory for our application; MMs are supplied by 
A7030DP 12-channel units, with maximum voltage of +3~kV and 
2~nA resolution of the current monitor; this resolution is
marginal respect to our needs and the off-set of the current
monitor is unstable. The power supply units are housed in two
main frame type SY 4527, one placed near the top detectors and 
one placed near the bottom ones to minimize the HV cable, which is  5~m. 
\par
HV channel setting within a detector is
highly correlated, making the correct manual operation 
a complex and risky task. Therefore, a custom-made 
control system has been  realized; it is written in 
C++ making use of the wxWidgets library 
and designed so to comply with the COMPASS 
experiment slow control system. The reference HV values are
accompanied by individual "OwnScale" of each channel to make 
fine-tuning for gain uniformity easier. An autodecrease HV 
algorithm is implemented to protect the detectors 
in case of a too high current spark rate. The voltage and 
current monitored by the power supply are logged at 1 Hz rate. 
The voltages are continuously corrected on the base of the
measured environmental parameters, namely temperature and 
pressure, in order to preserve the detector gain stability.
The correction algorithm is the result of dedicated 
laboratory tuning exercises:
\begin{equation}
V(P,T) = V_0  (1+0.5  (P/P_0 \cdot T_0/T -1 ))
\end{equation}
where V is the voltage, P is the absolute pressure in mbar, 
T is the temperature in degrees Kelvin 
and V$_0$, T$_0$ and P$_0$
refer to reference conditions.
\par
The control system is completed by a GUI interface 
for user access. 
\section{The electrical performance of the the novel MPGD-based photon detectors of COMPASS RICH-1}
No HV-level trip has been observed at the supply 
channels over about 
seven months of operation, thanks to the  schemes 
adopted for the HV distribution. Current sparks are present,
corresponding to temporary discharges. The stability of the 
voltage from the supply guarantees that only a portion of the 
detector is affected, namely, a physical segment in case of 
THGEMs and a pad area in case of MMs. The recovery time, 
monitored via the current information, is $\lesssim$10~s for 
current sparks in THGEM and about 1-2~s in MMs. Sparks 
in the two THGEMs are 100\% correlated. All the observed 
sparks in MMs are accompanied by sparks in THGEMs 
and not vice versa, suggesting that they are induced by misbehaviors in THGEMs. 
\par
\begin{figure}
\begin{center}
\includegraphics[width=0.4\textwidth]{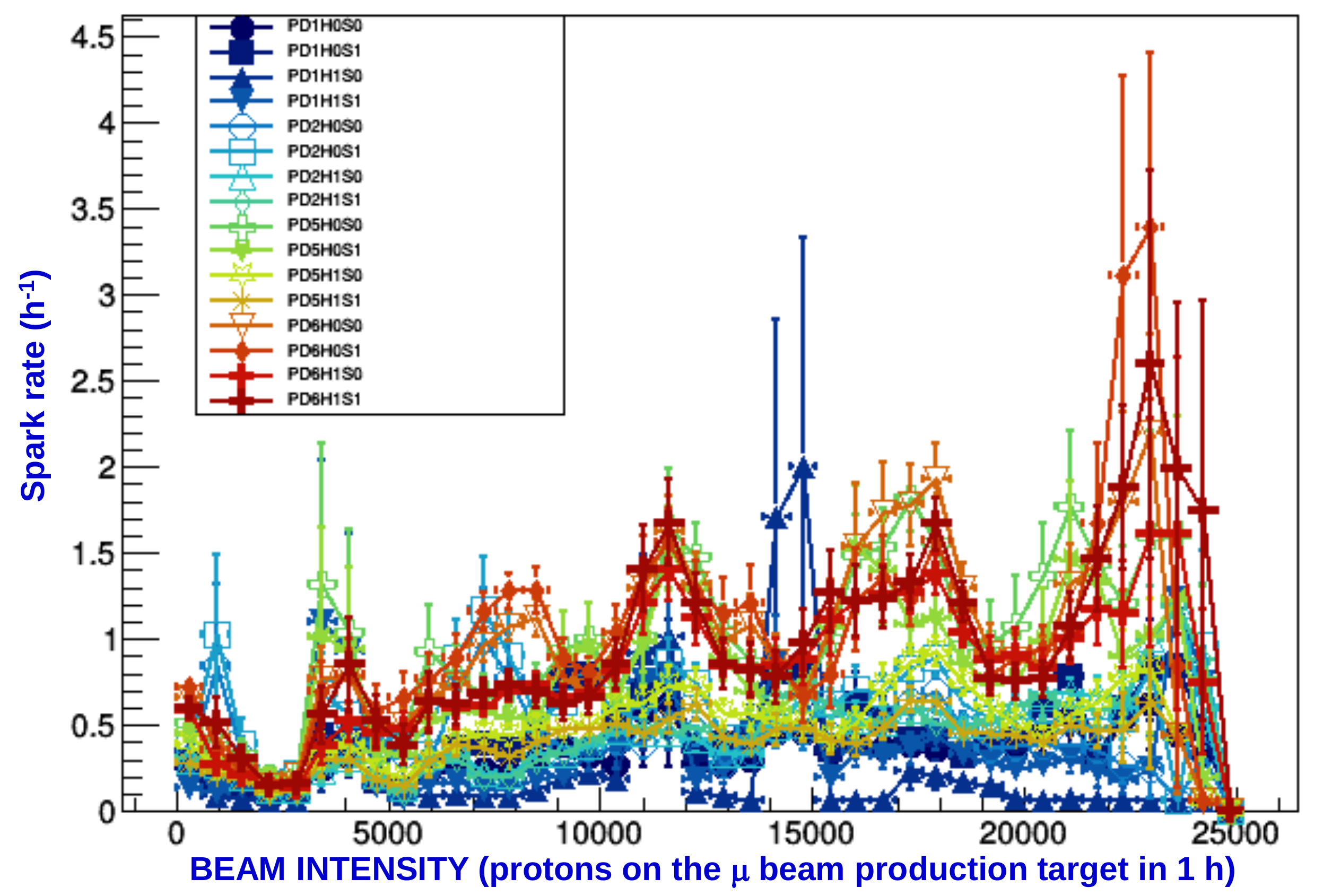}
\end{center}
\caption{\label{fig:spark-rate}
Spark rate versus beam intensity for the 16 sectors of the
four chambers.}
\end{figure}

Sparks rates, typically not exceeding 
20/h in total 
for the four novel detectors, 
do not exhibit a remarkable dependence on  
the beam intensity
(Fig.~\ref{fig:spark-rate}). Sparks correlated in time have 
been observed for THGEM segments, also when they are not 
contiguous. These observations suggest that a major source of
the observed sparks are cosmic rays that, occasionally, can 
produce relevant ionization in these detectors which are 
oriented almost vertically.
\par
Three sectors have exhibited weaker electrical stability 
during the first year of operation of the novel detector, 
corresponding to almost 20\% of the active surface.
The observations about the current spark correlations 
suggest that they are related to THGEM regions with weaker
performance. Therefore, 
a dedicated measurement campaign was performed in the 
2016-2017 Winter shut-down of the SPS, to identify, within 
each single sector, the weak segments in order to supply them 
separately and move the weak performance to smaller 
areas: a segment represent only 1\% of the total active area.
THGEM 1 and THGEM2 of each detector have been studied 
separately. Each segment of the THGEM under study was 
separately powered at increasing voltages, in a sequence of 
10-minute steps. The spark-rate was measured, segment by 
segment during the step and the voltage was decreased, instead 
of increasing it, if the number of sparks was exceeding 5.
The voltage and current parameters were logged and the 
exercise extended for a few consecutive days for each THGEM.
The total time that a segment could stand a given voltage, 
the number of sparks versus voltage and the spark rate at 
given voltage  were used to monitor the 
segment quality and they have provided consistent indications,
from which the weak segments have been identified.
A single weak segment was identified in the THGEMs used as THGEM 1, 
in spite of the presence of the CsI coating and, more 
in general, these THGEMs could stand higher 
voltage than the THGEMs used as THGEM 2. This unexpected outcome can be related
to the extra preparatory processes
required to use a THGEM as photocathode 
substrate: Ni-Au coating and one more cleaning step. 
In total 10 weak segments were identified and separately 
powered, corresponding to 10\% of the total active surface.
\par
A few local shorts have developed over months of operation in the MM gaps. They affect single, isolated pads, they are observed through an increase of the supply current and could not be removed by the application of reversed bias. We do not know the cause. After seven months of operation the observed rate is 1 per mil.
\par
The detector gain has been observed over 
periods of months by 
analyzing the collected data. The gain is extracted from the 
amplitude spectrum of the signals generated by the single 
photoelectrons; it is stable at about the 5\% level apart 
in a single sector, for which further investigation is ongoing.
\section{Conclusions}
In spite of the complexity of hybrid MPGDs,  the
HV system with sophisticated control, that we have 
implemented, guarantees safety operation, 
the collection of the
information required for understanding and monitoring 
the detector behavior and their satisfactory
electrical stability at gains around 15~k, a 
non-trivial 
achievement considering that, so far, 
all MPGDs operate in experiment at gain below 10~k.
\\
\par
\textbf{ACKNOWLEDGMENT}
\par
The activity is partially supported by the H2020 project
AIDA2020 GA no. 654168.

\end{document}